\documentclass{aa520} 

\begin{document}
  
\title{A toy model for coupling accretion disk oscillations to the
  neutron star spin.}

\author{J\'er\^ome P\'etri \inst{1}}

\offprints{J. P\'etri}

\institute {Max-Planck-Institut f\"ur Kernphysik, Saupfercheckweg 1,
    69117 Heidelberg, Germany.}

\date{Received 1 July 2005 / Accepted 2 September 2005}

\titlerunning{A toy model for accretion disk/neutron star coupling.}

\authorrunning{P\'etri}

\maketitle

\begin{abstract}

  Lee, Abramowicz \& Klu{\'z}niak~(\cite{Lee2004}) demonstrated
  numerically that rotation of neutron star couples with oscillations
  of its accretion disk, and excites resonances. No specific coupling
  was assumed, but magnetic field was suggested as the most likely
  one.  Following this idea, we show (P{\'e}tri~\cite{Petri2005a},
  paper~I) that if the neutron star is non-axially symmetric and
  rotating, its gravity may provide the coupling and excite
  resonances. Here, we return to the original suggestion that the
  coupling is of a magnetic origin, and demonstrate how does it work
  in terms of a simple, analytic toy-model.
  
\end{abstract}

\section{INTRODUCTION}

In some neutron star sources, the observed QPO frequencies obviously
depend on the neutron star spin. For example, difference in
frequencies of the double peaked QPO in the millisecond pulsar SAX
J1808.4-3658 is clearly equal to half of the pulsar spin (Wijnands et
al.~\cite{Wijnands2003}). This made Klu{\'z}niak et
al.~(\cite{Kluzniak2004}) to suggest that a resonance is excited by
coupling accretion disk oscillation modes to the neutron star spin.
The suggestion was fully confirmed by numerical simulations of the
coupling (Lee, Abramowicz \& Klu{\'z}niak, \cite{Lee2004}).  It was
found that a resonant response occurs when the difference between
frequencies of the two modes equals to one-half of the spin frequency
(as observed in SAX J1808.4-3658 and other "fast rotators"), and when
it equals to the spin frequency (as observed in "slow rotators" like
XTE J1807-294).
  
Lee et al.~(\cite{Lee2004}) suggested that the coupling is most likely
a magnetic one, but did not specified any concrete mechanism in their
numerical simulations, introducing the coupling by a purely formal
ansatz.  Following their idea, we discussed in terms of a simple
analytic toy-model (P{\'e}tri~\cite{Petri2005a}, Paper I) that also
(non-axially symmetric) gravitational field of rotating neutron star
may excite resonances in accretion disk oscillations.
   
Another point of view to account for the dichotomy between slow and
fast rotators is given by Lamb \& Miller~(\cite{Lamb2003}) who explain
it in the framework of the sonic-point beat frequency model.

In this Research Note, we discuss another toy model that provides the
coupling by the neutron star magnetic field. We use the same notation
as in Paper~I.

\section{THE MODEL}
\label{sec:Model}

In this section, we describe the main features of the model, starting
with a simple treatment of the accretion disk, assumed to be made of
non interacting charged single particles orbiting in the equatorial
plane of the star. Magnetohydrodynamical aspect of the disk such as
pressure and current are therefore neglected. Particles evolve in a
perfectly spherically symmetric gravitational potential. The asymmetry
arises from an eccentric misaligned dipolar magnetic field corotating
with the neutron star.

\subsection{Eccentric stellar magnetic field}

The periodically varying epicyclic frequencies are introduced by
adding a rotating asymmetric dipolar magnetic field to the background
gravity. Generally, when dealing with an oblique rotator, the location
of the magnetic moment~$\vec{\mu}$ generating the dipolar magnetic
field coincides with the centre of the neutron star (supposed to be a
perfect sphere).  In this paper, we lift this assumption and shift the
location of the magnetic moment to a point $\vec{r}_{\mathrm{s}}(t) =
(r_\mathrm{s}\ne0, \varphi_\mathrm{s} = \Omega_*\,t, z_\mathrm{s})$
inside the star such that
$||\vec{r}_{\mathrm{s}}(t)||=\sqrt{r_\mathrm{s}^2 + z_\mathrm{s}^2}\le
R_*$ where $R_*$ is the stellar radius. We use cylindrical coordinates
denoted by~$(r, \varphi, z)$.  Nevertheless, the origin of the
coordinate system coincides with the centre of the neutron star.
Furthermore, in order to compute analytically such kind of magnetic
field structure, we assume that the star is made of an homogeneous and
isotropic matter everywhere with total mass~$M_*$ and spinning around
its centre at an angular rate~$\vec{\Omega}_* = \Omega_* \,
\vec{e}_\mathrm{z}$, aligned with the z-axis. The magnetic field
induced by the dipolar source is therefore~:
\begin{equation}
  \label{eq:ChampMagBoiteux}
  \vec{B}(r,\varphi,z,t) = \frac{\mu_0}{4\,\pi} \left[ \frac{3 \,
      (\vec{\mu} \cdot \vec{R} ) \, \vec{R}}{R^5} - \frac{\vec{\mu}}{R^3} \right]
\end{equation}
The vector joining the source point $\vec{r}_{\mathrm{s}}(t) =
(r_\mathrm{s}, \varphi_\mathrm{s}, z_\mathrm{s})$ to the observer
point $\vec{r} = (r,\varphi,z)$ is~:
\begin{equation}
  \label{eq:VecPos}
  \vec{R}(t) = \vec{r} - \vec{r}_{\mathrm{s}}(t)
\end{equation}
Using the cylindrical frame of reference, the distance between source
point and observer is~:
\begin{equation}
  \label{eq:Distance}
  R^2 = r^2 + r_{\mathrm{s}}^2 - 2 \, r \, r_{\mathrm{s}} \cos\psi + ( z - z_{\mathrm{s}} )^2
\end{equation}
where the azimuth in the corotating frame is~$\psi = \varphi -
\Omega_* \, t$.  The magnetic moment anchored in the neutron star,
rotates at the stellar speed such that~:
\begin{equation}
  \label{eq:MomMag}
  \vec{\mu}(t) = \mu \, [ \sin\chi \, \{ \cos \, (\Omega_*\,t)
  \, \vec{e}_\mathrm{x} + \sin \, (\Omega_*\,t) \, \vec{e}_\mathrm{y} \} + \cos\chi \,
  \vec{e}_\mathrm{z} ]
\end{equation}
where the obliquity, i.e. the angle between $\vec{\mu}$ and
$\vec{\Omega}_*$, is denoted by $\chi$. Moreover, each component of
the magnetic field can be expressed as follows~:
\begin{eqnarray}
  \label{eq:ChampMagBoiteuxCylR}
  B_\mathrm{r} & = & \frac{\mu_0}{4\,\pi\,R^3} \, \left[ 
    \frac{3 \, (\vec{\mu} \cdot \vec{R}) \, ( r - r_\mathrm{s} \, \cos\psi) }{R^2}
    - \mu \, \sin\chi \, \cos\psi \right] \nonumber \\ \\
  \label{eq:ChampMagBoiteuxCylP}
  B_\varphi & = & \frac{\mu_0}{4\,\pi\,R^3} \, \left[ 
    \frac{3 \, (\vec{\mu} \cdot \vec{R}) \, r_\mathrm{s} \, \sin\psi }{R^2}
    + \mu \, \sin\chi \, \sin\psi \right] \\
  \label{eq:ChampMagBoiteuxCylZ}
  B_\mathrm{z} & = & \frac{\mu_0}{4\,\pi\,R^3} \, \left[ 
    \frac{3 \, (\vec{\mu} \cdot \vec{R}) \, ( z - z_\mathrm{s} ) }{R^2}
    - \mu \, \cos\chi \right] \\
  \vec{\mu} \cdot \vec{R} & = & \mu \, \left[ \sin\chi \, ( r \, \cos\psi - r_\mathrm{s} )
    + \cos\chi \, ( z - z_\mathrm{s} ) \right]
\end{eqnarray}
The total linear response of the disk is then the sum of each
perturbation corresponding to one particular azimuthal mode~$m$.
Because the perturber is inside the star and the disk never reaches
the stellar surface, the Fourier coefficients of each
component~$B_i^m$ ($i={\mathrm{r},\varphi,\mathrm{z}}$) never diverge.
It is convenient to introduce the Fourier decomposition of the
magnetic field component
Eq.~(\ref{eq:ChampMagBoiteuxCylR})-(\ref{eq:ChampMagBoiteuxCylZ}) in
order to describe the response of the test particle. Moreover, because
the evaluation of the Fourier coefficients requires to integrate terms
containing $\cos\,(m\,\psi)$ in the integrand, the value of these
coefficients decreases rapidly with increasing azimuthal number~$m$.
As a result, only low azimuthal modes will influence significantly the
evolution of the disk. Keeping only the few first terms in the
expansion is sufficient to achieve reasonable accuracy.  For
discussing the results, we only keep the three first modes, namely,
the dipolar, quadrupolar and octupolar moments ($m=1,2,3$
respectively).

\subsection{Equation of motion for a charged test particle}

All particles evolve in the gravitatomagnetic field imposed by the
rotating neutron star. To keep things as simple as possible, their
motion is described in the guiding centre approximation. The drift
arising from the gyration around the local magnetic field is not an
essential feature we want to discuss here. As a consequence, magnetic
curvature and gradient as well as gravity drift motions are ignored in
this study. Nevertheless, the main characteristic consisting of a
periodic variation in the epicyclic frequencies is preserved. The
equation of motion then reads~:
\begin{equation}
  \label{eq:EqMvt}
  \ddot{\vec{G}} = \vec{g} + \frac{q_\mathrm{e}}{m_\mathrm{e}} \, \dot{\vec{G}} \wedge \vec{B}
\end{equation}
where $\vec{G}$ is the location of the guiding centre and the dot
means time derivative $d/dt$. The gravitational field of the
star~$M_*$ is denoted by~$\vec{g} = \vec{\nabla} ( G \, M_* /
\sqrt{r^2+z^2})$. The mass and the charge of the particle are denoted
respectively by $m_\mathrm{e}$ and $q_\mathrm{e}$. If for instance the
magnetic gradient drift is taken into account, the term $- (
\mu_\mathrm{e} / 2 \, m_\mathrm{e} \, B ) \, \vec{\nabla} B^2$ should
be added to the right hand side of Eq.~(\ref{eq:EqMvt}) where
$\mu_\mathrm{e}$ is an adiabatic invariant, namely the magnetic moment
of the test particle gyrating along the local field line. This would
introduce another modulation of the epicyclic frequencies which is
already included in the Lorentz force $q_\mathrm{e} \, \dot{\vec{G}}
\wedge \vec{B}$.  Thus, the physical behaviour is not changed by
neglecting the drift motion of the guiding centre $\vec{G}$. Expressed
in cylindrical coordinates, Eq.~(\ref{eq:EqMvt}) develops into
\begin{eqnarray}
  \label{eq:EquationMvtRad}
  \ddot{r} - r \, \dot{\varphi}^2 & = & g_\mathrm{r} + \frac{q_\mathrm{e}}{m_\mathrm{e}} \, 
  ( r \, \dot{\varphi} \, B_\mathrm{z} - \dot{z} \, B_\varphi ) \\
  \label{eq:EquationMvtAzi}
  2 \, \dot{r} \, \dot{\varphi} + r \, \ddot{\varphi} & = & \frac{q_\mathrm{e}}{m_\mathrm{e}} \, 
  ( \dot{z} \, B_\mathrm{r} - \dot{r} \, B_\mathrm{z} ) \\
  \label{eq:EquationMvtVer}
  \ddot{z} & = & g_\mathrm{z} + \frac{q_\mathrm{e}}{m_\mathrm{e}} \, 
  ( \dot{r} \, B_\varphi - r \, \dot{\varphi} \, B_\mathrm{r} )
\end{eqnarray}
The magnetic field $\vec{B}$ is assumed to be weak enough for the flow
to remain essentially hydrodynamical (weakly magnetized thin disk
approximation). We therefore treat $\vec{B}$ as a perturbation of
order $\varepsilon\ll1$. The perturbations induced in the flow are of
the same order of magnitude than $\vec{B}$, i.e. of order
$\varepsilon$. The perturbed orbit and velocity of the test particle
in the radial and vertical direction are also of order $\varepsilon$,
$\{\dot{r},\dot{z}\} = \mathrm{O}(\varepsilon)$ whereas the azimuth
varies as $\dot{\varphi} \approx \Omega_\mathrm{k} \approx 
\sqrt{G\,M_*/r_0^3} + \mathrm{O}(\varepsilon)$, the Keplerian orbital
frequency at the radius of the orbit $r_0$.  In the equations of
motion (\ref{eq:EquationMvtRad}), (\ref{eq:EquationMvtAzi}) and
(\ref{eq:EquationMvtVer}), terms such as $\{\dot{r}, \dot{z}\}
\times\,B_i$ with $i=\{\mathrm{r}, \varphi, \mathrm{z}\}$ are second
order $\mathrm{O}(\varepsilon^2)$ and we neglect them.  According to
this simplification, the right hand side of
Eq.~(\ref{eq:EquationMvtAzi}) vanishes.  Eq.~(\ref{eq:EquationMvtAzi})
states the conservation of angular momentum of the particle and
integrates into $L = m\,r^2\,\dot{\varphi} = \mathrm{const}$ where $L$
is the angular momentum of the particle.  This is an obvious integral
of motion for this problem (to the aforementioned approximation).

However, we are only interested in the vertical motion experienced by
the test particles in response to the perturbed magnetic field.
Indeed, the response of an accretion disc to an inclined rotating
magnetic dipole has been studied by Terquem \&
Papaloizou~(\cite{Terquem2000}).  They showed that the vertical
displacement (i.e. the warping) is the dominant effect in the thin
disc while the horizontal perturbations can be neglected.  Perturbing
(\ref{eq:EquationMvtVer}) and developing to first order in the
perturbation around the equilibrium Keplerian orbit (contained in the
equatorial plane) defined by $(r_0, \varphi_0 = \Omega_\mathrm{k}\,t,
z_0=0)$, the vertical motion reads~:
\begin{equation}
  \label{eq:EquationMvtVer1}
  \ddot{z} = g_\mathrm{z} - \frac{q_\mathrm{e}}{m_\mathrm{e}} \, r_0 \, \Omega_\mathrm{k} \, B_\mathrm{r}
\end{equation}
To avoid mathematical irrelevant complications, we assume the rotator
to be ``aligned'' with the star in the sense that $\vec{\mu}$ and
$\vec{\Omega}_*$ are parallel ($\chi=0$). Therefore the radial component of the
magnetic field reads~:
\begin{equation}
  \label{eq:Br}
  B_\mathrm{r} = \frac{3 \, \mu_0 \, \mu}{4\,\pi} \, \frac{( z - z_\mathrm{s} ) 
    \, ( r - r_\mathrm{s} \, \cos\,\psi )}{R^5}
\end{equation}
By developing in a Fourier series, we obtain~:
\begin{equation}
  \label{eq:FourierBr}
  B_\mathrm{r} = ( z - z_\mathrm{s} ) \, \sum_{m=0}^{+\infty} 
  B_\mathrm{r}^m(r,z) \, \cos \, (m\,\psi)
\end{equation}
where the Fourier coefficients are given by~:
\begin{equation}
  \label{eq:FourierCoeff}
  B_\mathrm{r}^m(r,z) = \frac{3\,\mu_0 \, \mu}{4\,\pi} \, \frac{2-\delta_m^0}{2\,\pi} \, \int_0^{2\pi} 
  \frac{ r - r_\mathrm{s} \, \cos\,\psi}{R^5} \, \cos \, (m\,\psi) \, d\psi
\end{equation}
where $\delta_m^0$ is the Kronecker symbol. Note that $B_\mathrm{r}^m$
do not have the dimension of a magnetic field because of their
definition Eq.~(\ref{eq:FourierBr}). These coefficients, which are
function of the space position $(r,z)$ decrease with increasing
azimuthal number~$m$ (for $(r,z)$ fixed).  Keeping the first few
coefficients is sufficient to achieve a reasonable accuracy (we retain
the first three). Putting the expansion Eq.~(\ref{eq:FourierBr}) into
the vertical equation of motion Eq.~(\ref{eq:EquationMvtVer1}), we get
the fundamental equation to describe vertical forced oscillations of a
test particle as follows~:
\begin{eqnarray}
  \label{eq:Oscillation}
  \left [ \Omega_\mathrm{k}^2 + \frac{q_\mathrm{e}}{m_\mathrm{e}} \, r_0 \, \Omega_\mathrm{k} 
    \, \sum_{m=0}^{+\infty} B_r^m(r_0,z_0) \, 
    \cos \, \{ m(\Omega_\mathrm{k}-\Omega_*) \, t \} \right ] \, z & & 
  \nonumber \\
  + \ddot{z} = 
  \frac{q_\mathrm{e}}{m_\mathrm{e}} \, r_0 \, \Omega_\mathrm{k} \, z_\mathrm{s} \, \sum_{m=0}^{+\infty} 
  B_r^m(r_0,z_0) \, \cos \, \{ m \, (\Omega_\mathrm{k}-\Omega_*) \, t \} \nonumber \\
\end{eqnarray}
Note that the Fourier coefficients $B_r^m(r_0,z_0)$ in this last
equation are evaluated at the location of the unperturbed orbit and do
no longer depend on the perturbed position $(r,z)$. To the lowest
order of the expansion, this approximation is justified. We recognize
a Hill equation (periodic variation of the eigenfrequency of the
system on the left hand side) with a periodic driving force (on the
right hand side).

\subsection{Resonance conditions}

This equation is very similar to the one obtained in the case where
solely gravity perturbation exists. The discussion is therefore
exactly the same as in paper~I. Here we recall the main results,
adapted to the magnetic configuration.  Eq.~(\ref{eq:Oscillation})
describes an harmonic oscillator with periodically varying
eigenfrequency which is also excited by a driven force.  It is well
known that some resonances will therefore occurs in this system.
Namely, we expect three kind of resonances corresponding to~:
\begin{itemize}
\item a {\it corotation resonance} at the radius where the angular
  velocity of the test particle equals the rotation speed of the
  magnetic structure (which is the equal to the stellar rotation
  rate). Corotation is only possible for prograde motion. The
  resonance condition determining the corotating radius is simply
  $\Omega_\mathrm{k} = \Omega_* $;
\item a {\it driven resonance} at the radius where the vertical
  epicyclic frequency equals the frequency of each mode of the
  magnetic perturbation as seen in the locally corotating frame. The
  resonance condition is $m \, | \Omega_* - \Omega_\mathrm{k} | =
  \kappa_\mathrm{z}$;
\item a {\it parametric resonance} related to the time-varying
  vertical epicyclic frequency, (Hill equation). The rotation of the
  magnetosphere induces a sinusoidally variation of the vertical
  epicyclic frequency leading to the well known Mathieu's equation for
  a given azimuthal mode $m$. The resonance condition is derived as
  followed~:
  \begin{equation}
    \label{eq:ResPara}
    m \, |\Omega_* - \Omega_\mathrm{k}| = 2 \, \frac{\kappa_\mathrm{z}}{n}
  \end{equation}
\end{itemize}
where $n\ge 1$ is a natural integer.  Note that the driven resonance
is a special case of the the parametric resonance for~$n=2$. However,
their growth rate differ by the timescale of the amplitude
magnification.  Driving causes a linear growth in time while
parametric resonance causes an exponential growth. We also rewrite the
vertical epicyclic frequency as~$\kappa_\mathrm{z}$ instead of
$\Omega_\mathrm{k}$ in order to apply the results to a more general
case which could include stronger magnetic fields or general
relativistic effects.

Consequently, the resonance conditions for the magnetized rotator are
exactly the same as for the unmagnetized rotator of paper~I, as long
as the oscillations remain in the linear regime (i.e. in the thin disk
approximation for which the vertical motion of the particle remains
small with respect to the radius of the unperturbed orbit). For more
details and a discussion on these results, we refer the reader to
paper~I. 

The dichotomy between the QPOs in fast and slow rotators has been
explained by Lee, Abramowicz \& Klu{\'z}niak (2004) as a consequence
of the fact that the coupling between the neutron star spin and the
modes of accretion disk oscillations excites possible resonances at
different locations in the disk, either very close to the star, or far
away from it. While we agree with this results, we point out that (as
found in Paper I) the spectrum of modes that {\it could be in a
  resonance} is more complex than that consider by Lee et al.  Their
discussion was concentrated on the 3:2 resonance that occurs between
the radial and vertical epicyclic modes {\it only in strong gravity}.
We identified possible forced resonances between different modes, that
may occur in both strong and weak gravity. Our analysis was done in a
linear regime, so, at the moment, we are only able to say that such
resonances may in principle exist\footnote{I was informed by
  M.~Abramowicz, the referee, that unpublished numerical results of
  Lee et al. fully confirms existence of these additional, weak
  gravity's, resonances.}.

\section{CONCLUSION}
\label{sec:Conclusion}

The toy-model discussed in this {\it Research Note} illustrates the
idea (Klu{\'z}niak et al., \cite{Kluzniak2004}; Lee et al.
\cite{Lee2004}; P{\'e}tri \cite{Petri2005a}) that the double peak QPOs
in neutron stars sources may be due to coupling rotation of neutron
star sources to modes in accretion disk oscillations, and exciting
resonances.
  
The model is physically very specific. It explicitly shows how the
rotating magnetic field of the neutron stars could couple with the
disk dynamics and oscillations.  It confirms general discussion and
numerical results of Klu{\'z}niak et al., (\cite{Kluzniak2004}) and
Lee et al.  (\cite{Lee2004}), in particular the important one that the
strongest resonant response occurs when the difference between
frequencies of the two modes equals to one-half of the spin frequency
(as observed in SAX J1808.4-3658 and other "fast rotators"), and when
it equals to the spin frequency (as observed in "slow rotators" like
XTE J1807-294).

When the MHD nature of the flow is taken into account, QPOs can be
explained by a mechanism similar to those exposed here
(P\'etri~\cite{Petri2005b}). However, in an accreting system in which
the neutron star is an oblique rotator, we expect a perturbation in
the magnetic field to the same order of magnitude than the unperturbed
one. Therefore, the linear analysis developed in this paper has to be
extended to oscillations having non negligible amplitude compared to
the stationary state. Nonlinear oscillations therefore arise naturally
in the magnetized accretion disk. Abramowicz et al.
(\cite{Abramowicz2003}) showed that the non-linear resonance for the
geodesic motion of a test particle can lead to the 3:2 ratio for the
two main resonances. Nevertheless, an extension to 3D MHD flow in
curved spacetime is required to make quantitative accurate predictions
of the peak frequencies variation correlated with the accretion rate
for instance.

\begin{acknowledgements}
  I am grateful to the referee Marek A. Abramowicz for his valuable
  comments and remarks.  This work was supported by a grant from the
  G.I.F., the German-Israeli Foundation for Scientific Research and
  Development.
\end{acknowledgements}


\begin{thebibliography}{2}
  
\bibitem[2003]{Abramowicz2003}
  Abramowicz, M.~A., Karas, V., Klu{\'z}niak, W. ,Lee, W.~H. \& Rebusco, P.,
  2003, \pasj, 55, 467
  
%\bibitem[2004a]{Kluzniak2004a}
%  Klu{\' z}niak, W., Abramowicz, M.~A., Kato, S., Lee, W.~H. \& Stergioulas, N.,
%  2004a, \apjl, 603, L89

\bibitem[2004]{Kluzniak2004}
  Klu{\' z}niak, W., Abramowicz, M.~A. \& Lee, W.~H., 
  AIP Conf. Proc. 714: X-ray Timing 2003: Rossi and Beyond, 2004, 379

\bibitem[2003]{Lamb2003}
  Lamb, F.~K. \& Miller, M.~C., 2003, astro-ph/0308179

\bibitem[2004]{Lee2004}
  Lee, W.~H., Abramowicz, M.~A., \& Klu{\' z}niak, W., 2004, \apjl, 603, L93

\bibitem[2005a]{Petri2005a}
  P\'etri J., 2005a, \aap, 439, L27, paper~I

\bibitem[2005b]{Petri2005b}
  P\'etri J., 2005b, \aap, 439, 443

\bibitem[2000]{Terquem2000}
  Terquem, C. \& Papaloizou, J.~C.~B., 2000, \aap, 360, 1031

\bibitem[2003]{Wijnands2003} 
  Wijnands, R., van der Klis, M., Homan, J., Chakrabarty, D., 
  Markwardt, C.~B. \& Morgan, E.~H., 2003, \nat, 424, 44

\end{thebibliography}
\end{document}